\documentstyle[epsfig]{article}

\def\gsim{ \lower .75ex \hbox{$\sim$} \llap{\raise .27ex \hbox{$>$}} }
\def\lsim{ \lower .75ex\hbox{$\sim$} \llap{\raise .27ex \hbox{$<$}} }

\def\sc{Schwarzschild}

\begin{document}
\begin{center}
{\huge \bf Extragalactic relativistic jets}
\vskip 0.3 cm
{\large Gabriele Ghisellini}
\vskip 0.3 cm

{\it INAF -- Osservatorio Astronomico di Brera}

\end{center}

\vskip 0.3 cm
\noindent
{\large \bf Abstract}

\normalsize
\noindent
Extragalactic relativistic jets are engines able to carry out to large distances
a huge amount of power, not only in the form of radiation, but especially in the
form of kinetic energy of matter and fields.
As such, they can be thought as one of the most efficient engines of Nature,
perhaps even more efficient than accretion.
We are starting to disclose these features through a detailed study of their
properties, made possible by the analysis of the energy band where they emit
most of their electromagnetic output, namely the $\gamma$--ray band.
That is why the observations by the {\it Fermi} satellite and by the ground based 
Cherenkov telescopes are crucial to understand extragalactic jets.
At the start, we believe they are magnetically dominated. And yet, on the scale 
where they emit most of their luminosity, their power is already in the form
of kinetic energy of particles.
The spectral properties of bright sources show a trend, controlled mainly by 
the bolometric apparent luminosity.
With improved sensitivity, and the detection of weaker sources, we can explore
the idea that the spectral trends are a result of the same physical quantities
controlling the emission of non--jetted sources: the black hole mass and the 
accretion rate. 
This is based on recent results on sources
showing a thermal component in their spectrum, besides a non--thermal continuum.
That the jet power should be linked to accretion is intriguing. 
Most of the apparent diversity of extragalactic radio sources can then be
understood on the basis of the viewing angle, controlling the relativistic
Doppler boosting of the emission, the black hole mass and the accretion rate.  

{\bf Keywords:} 95.85.Pw, 98.54.Cm, 98.62.Mw, 98.62.Nx

{gamma--ray, blazars, accretion disks, jets }


\section{Introduction}
We now know that almost every galaxy hosts a supermassive black hole, from
millions to billions of solar masses.
The vast majority of them is ``silent", but about 1\% of them accretes enough
matter to become visible and overtake the emission from the ensemble of stars
in the entire galaxy.
About 10\% of these systems, thus 0.1\% of the supermassive black holes, is able 
to launch relativistic jets.
These jets design spectacular structures in the radio, with sizes far exceeding
the dimensions of the host galaxy, reaching in a few cases the megaparsec scale.
The emitting matter is moving relativistically, with bulk Lorentz factors
$\Gamma > 10$ (i.e. $\beta>0.995$), therefore boosting the emission in 
a cone of semi--aperture $1/\Gamma<5^\circ$. 
Objects seen face on are dramatically different from objects observed from the side. 
The former are called ``blazars" (BL Lac objects and flat spectrum radio 
quasars, or FSRQs for short).
The latter are radio--galaxies.
Because of the relativistic boosting, blazars are powerful,  
active, rapidly variable and detectable from the farthest distance.

Although first discovered in the radio, jets do not emit much in this band.
They instead emit most of their radiation at the other extreme of the electromagnetic 
spectrum, namely in the $\gamma$--ray band.
This is why we had to wait the launch of the {\it Compton Gamma Ray Observatory, CGRO}
and its EGRET instrument [0.1--30 GeV] to discover that all blazars were strong $\gamma$--ray
emitters as a class (Hartman et al. 1999; Nandikotkur et al. 2007).
At last we could know the bolometric luminosity and the entire spectral energy distribution.
This turned out to be double humped.
The first hump is interpreted as synchrotron emission. 
The second, high energy, one is usually interpreted as inverse Compton (IC)
emission (but there are other suggestions, see Mannheim 1993; M\"ucke et al. 2003; 
B\"ottcher 2007; Aharonian 2000; M\"ucke \& Protheroe 2001).

\begin{figure}
\vskip -0.8 cm
\begin{tabular}{l l}
\includegraphics[height=0.34\textheight]{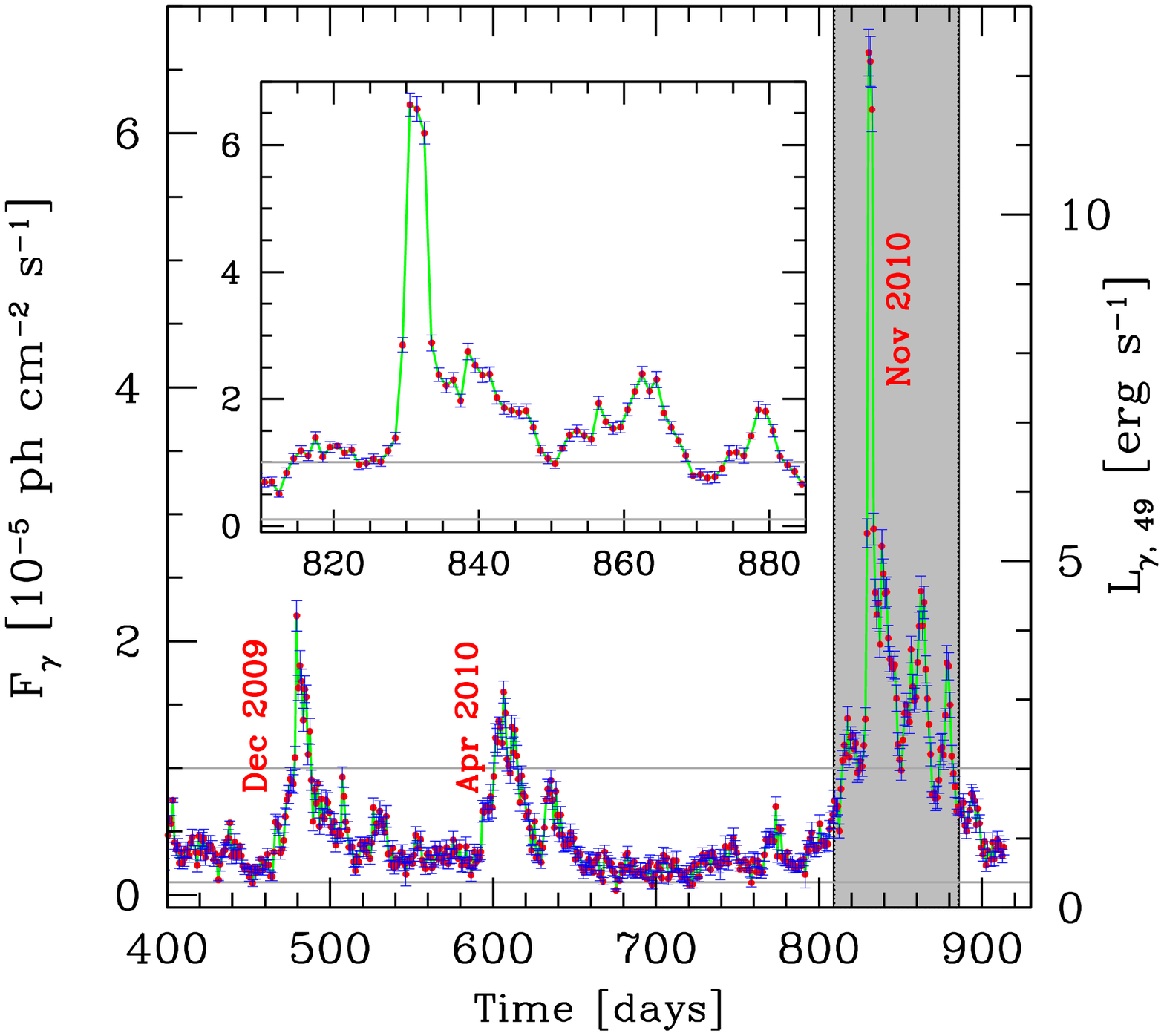}
& \includegraphics[height=0.34\textheight]{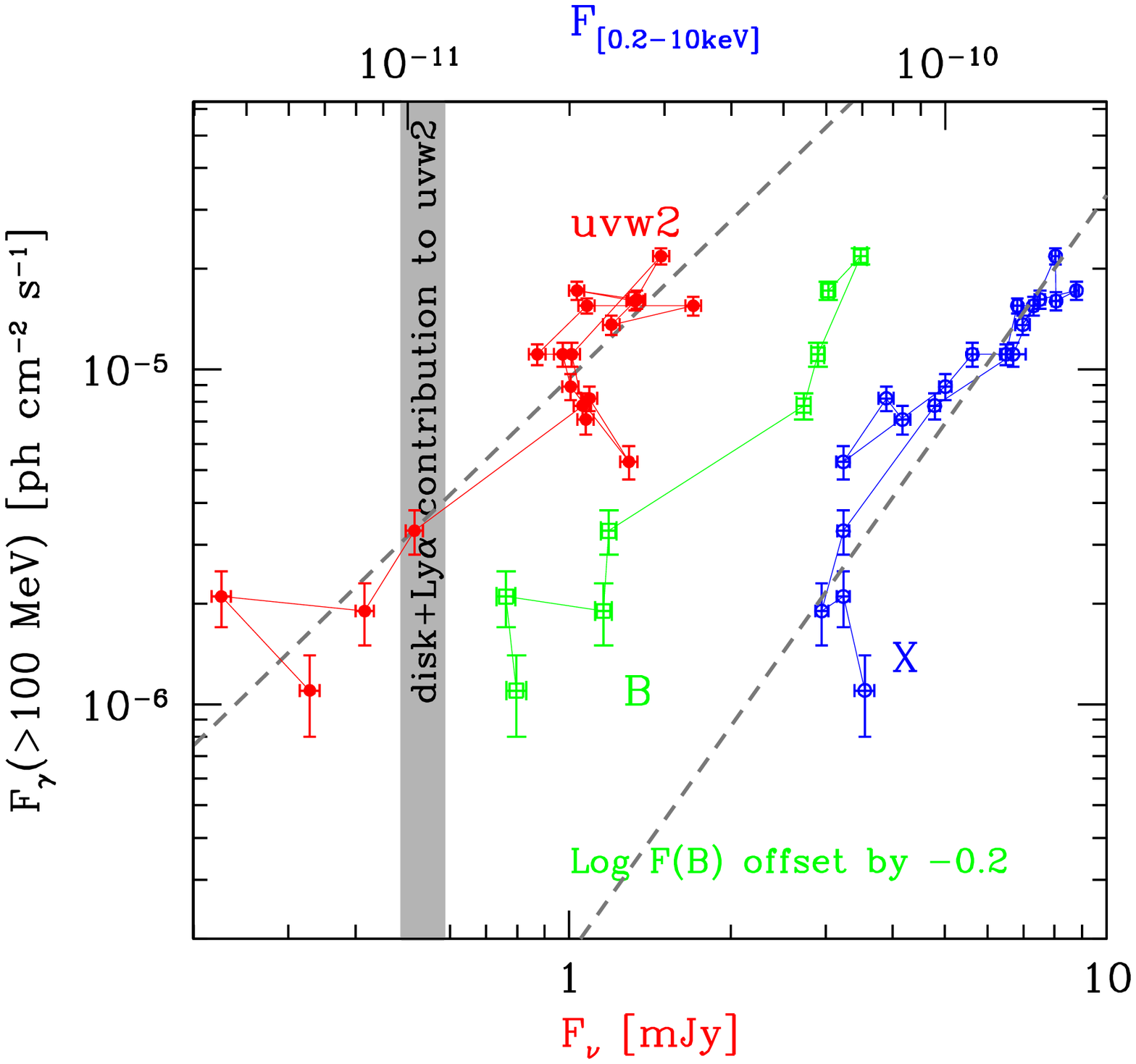}
\end{tabular}
\vskip -0.5 cm
\caption{
{\bf Left:}
Light curve of 3C 454.3 in the $\gamma$--ray band of {\it Fermi}/LAT showing three major flares.
Time bins are 1 day.
The inset is a zoom of the last one, occurred in November 2010.
Time bins are 1 day long.
Assuming that the 0.1--10 GeV spectrum is a power law of energy index $\alpha_\gamma\sim 1.2$, the
we can calculate the corresponding luminosity (assuming isotropy), that exceeded $10^{50}$ erg s$^{-1}$
in Nov. 2010.
Note also the fast variability. 
A detailed analysis showed a variability timescale (i.e. the time to halve or double the flux)
of a few hours (Tavecchio et al. 2010; Foschini et al. 2011).
{\bf Right:}
Correlation between the $\gamma$--ray, the X--ray and the UV flux of 3C 454.3 during the
Dec. 2009 flare (from Bonnoli et al. 2011).
}
\label{454}
\end{figure}

The apparent $\gamma$--ray luminosity (calculated assuming isotropy) can be huge, 
exceeding for a few days $10^{50}$ erg s$^{-1}$ (left panel and right y--axis of 
Fig. \ref{454}; see also Abdo et al. 2011).
Coordinated variability of the flux at different frequencies (belonging to
the two humps; Bonnoli et al. 2011; see the right panel of Fig. \ref{454}) implies two important facts: 
i) the same population of leptons is responsible for both humps and therefore 
ii) there must be a single emitting zone.


\begin{figure}
\vskip -1 cm
\includegraphics[height=0.73\textheight]{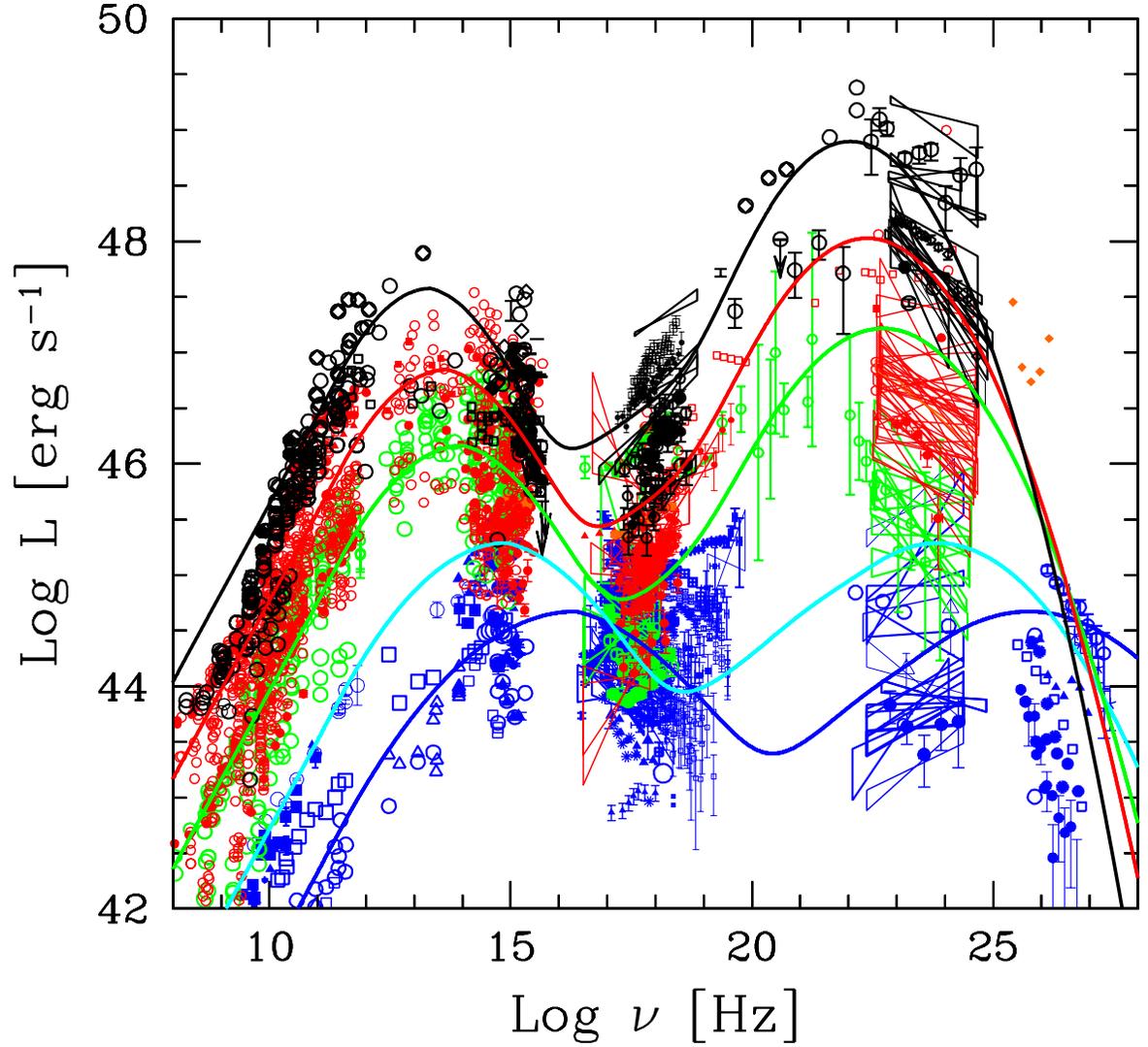}
\vskip -0.5 cm
\caption{
The blazar sequence.
Solid lines are the phenomenological SED supposed to describe the SED of
blazar of different bolometric luminosities, according to Fossati et al. (1998)
and Donato et al. (2001).
The data are the SED of real sources detected in the first 3 months of
survey by the {\it Fermi} satellite (LBAS catalog: Abdo et al. 2009), 
divided (different colors) into
bins of different $\gamma$--ray luminosities in the [0.1--10 GeV] range:
$\log L_\gamma<45.5$ (blue), $45.5<\log L_\gamma<46.5$ (green)
$46.5<\log L_\gamma<47.5$ (red), $\log L_\gamma>48.5$ (black).
}
\label{sequence}
\end{figure}

There is still some discussion about the origin of the seed photons that are scattered
at higher energies:
these could be the synchrotron photons produced internally to the jet (synchrotron self--Compton
process, SSC; Maraschi, Ghisellini \& Celotti 1992; Bloom \& Marscher 1996), or else photons
produced outside the jet (and in this case the process is called External Compton, EC).
In fact, at least in powerful blazars, there are several important photon sources:
radiation coming directly from the accretion disk (Dermer \& Shlickeiser 1993), or re--isotropized
by the clouds producing the broad emission lines (Sikora, Begelman \& Rees 1994);
IR radiation produced by a dusty torus surrounding the accretion disk (and extending
to a few parsecs from it; Sikora et al. 2002).

At some level, all these photon sources contribute, with a relative weight
that is dependent upon the distance of the emitting region from the 
black hole (Ghisellini \& Tavecchio 2009). 
But since BL Lac objects have weak or absent emission lines, the EC process
in these objects should be less important.
On the contrary, FSRQs with broad emission lines similar to radio--quiet
objects (of similar disk luminosity) should have more seed photons to scatter,
and thus produce a relatively stronger high energy hump.
Although some exceptions exist, this is what we see, and this is at the core
of the explanation of the so called ``blazar sequence": a well defined
trend between the overall spectral shape (or spectral energy distribution, SED)
and the apparent bolometric luminosity of blazars (Ghisellini et al. 1998).
Fig. \ref{sequence} shows that at low luminosities the SED of blazars 
(mostly BL Lac) peaks at large frequencies (synchrotron hump: opt--UV--soft X--ray,
high energy hump: GeV--TeV). 
The two humps have similar luminosities. 
Increasing the bolometric power, the SED shifts to smaller frequencies, and
the high energy hump prevails.
Fig. \ref{sequence} shows, as lines, the phenomenological SEDs originally
derived by Fossati et al. (1998) and Donato et al. (2001) on the basis of
complete sample of blazars selected at radio or X--ray frequencies.
Blazars detected in the $\gamma$--ray range were vey few.
It is interesting therefore to compare these early representation of the
blazar sequence with the bright blazars detected by {\it Fermi} in the
3 months all sky survey (LBAS catalog: Abdo et al. 2009), 
divided into $\gamma$--ray luminosity bins.
We can see that the solid lines represent rather well the average SEDs,
but tend to over--predict the high energy flux at intermediate $\gamma$--ray
luminosities (red and green lines).
This can be due to the fact that {\it Fermi}, $\sim$20 times more
sensitive than EGRET, starts to observe blazars not only in their
high states, but also when they are not extremely active.

\section{TeV BL Lacs}

At the low luminosity end of the blazar sequence we find 
BL Lacs whose synchrotron spectrum peaks in the UV or soft
X--rays (and sometimes even in the hard X--ray band, as in Mkn 501;
Pian et al. 1998), and the high energy hump extends to the TeV band.
In the {\it Fermi} band they are relatively weak, but hard
(i.e. energy spectral indices $\alpha_\gamma<1$).
The interest of these objects\footnote{see http://www.mpp.mpg.de/$\sim$rwagner/sources/
for an updates list of TeV detected blazars.
} 
concerns three main aspects:
\begin{itemize}
\item 
Since the emitting electrons reach TeV energies, they
give information about the acceleration mechanism at its extreme.
Particularly challenging in this respect are the ultrafast variations
seen in PKS 2155--304 (Aharonian et al. 2007) and in Mkn 501 (Albert et al. 2007),
showing a doubling of the TeV flux in about 3--5 minutes.
The causality argument, implying $R<ct_{\rm var}\delta$ where $\delta$ is the Doppler factor
(Begelman, Fabian \& Rees 2008),  
means that such small $t_{\rm var}$ cannot be any longer indicative of 
the size of the black hole (as instead is in the 
``internal shock scenario" (Sikora et al. 1994; Ghisellini 1999; 
Spada et al. 2001; Guetta et al. 2004).
The proposed solutions suggest that the emitting region responsible
for the ultrafast variations has a dimension much smaller than the
cross sectional radius of the jet, moving with a very large bulk Lorentz
factors. 
This can be produced by small reconnecting regions 
(Giannios et al. 2009a; ``jet in the jet" model) or ``needles" of 
emitting leptons (Ghisellini et al. 2009) very aligned with the line of sight.

\item 
Since the TeV photons can be absorbed by the $\gamma$--$\gamma\to e^\pm$ 
process interacting with the cosmic IR background, they can give
information about the latter (e.g. Mazin \& Raue 2007).
Using GeV (i.e. {\it Fermi}) and TeV (i.e. Cerenkov telescopes) data
greatly helps to estimate the amount of the absorption (important above 
a few hundred GeV). 

\item
The electron--positron pairs created by the above process have $\sim$TeV energies
and can scatter photons of the cosmic microwave background (CMB) up to
$\sim$GeV energies.
In the absence of any intergalactic magnetic field $B$ this reprocessed flux
much be equal to the absorbed one.
However, if a magnetic field is present, the pairs start to gyrate while
cooling, spreading their reprocessed flux in a broader beaming cone.
As a result, the received reprocessed flux is less than in the $B=0$ case.
Therefore the knowledge of the absorbed TeV flux together with the 
GeV one gives an indication of $B$. 
Since the received GeV flux can be the sum of the reprocessed radiation and 
the flux intrinsically produced by the source, what is derived is a lower 
limit on $B$ (Neronov \& Vovk ; Tavecchio 2010; see Dermer et al. 2011
and Tavecchio et al. 2011 for some caveats about the use of this method).

\end{itemize}

\section{The Blazars' divide}

The blazar sequence predicts that the high energy hump peaks 
at a frequency that shifts to smaller values by increasing the
bolometric apparent luminosity.
This is a simplification for several reasons.
First, the sources do vary, especially in the $\gamma$--ray band,
by huge factors (i.e. even more that two orders of magnitude).
Second, the use of the apparent luminosity depends on the level of Doppler boosting 
(therefore on the viewing angle). 
Sources very powerful when viewed on axis becomes much fainter at larger viewing angles 
$\theta_{\rm v}$ (i.e. there is a factor $\sim 16$ going from $\theta_{\rm v}=0$
to $\theta_{\rm v}=1/\Gamma$). 
Therefore a blazar that appears powerful on axis becomes 
much weaker if seen --say -- at 10$^\circ$, with a high energy peak
that is shifted to smaller values (e.g. contrary to what 
naively expected from the blazar sequence).
Note that these objects {\it must} exist.
Third, blazars with different black hole masses are expected to have different
jet luminosities (Ghisellini \& Tavecchio 2008) but an overall SED controlled
by the Eddington ratio (e.g. for smaller black hole masses the jet luminosity
could be smaller, but the two humps should peak at small frequencies).

Given all these caveats, one may wonder why the blazar sequence does work at all.
Most likely, the answer is in the fact that we are still detecting, in the
$\gamma$--ray band, the tip of the iceberg of a much broader distribution
of properties.
In $\gamma$--rays we still detect the brightest sources, that are the most aligned and 
with a greater black hole mass.

This explains why, for the brightest {\it Fermi} blazars detected at more than $10\sigma$ confidence 
level in the first three months of operations, there is a relation between $\alpha_\gamma$ and $L_\gamma$,
with a relatively well defined ``divide" in luminosity ($L_\gamma\sim 10^{47}$erg s$^{-1}$)
between BL Lacs and FSRQs (see Fig. 1 in Ghisellini, Maraschi \& Tavecchio 2009).
The most luminous FSRQs reach $L_\gamma\sim 10^{49}$erg s$^{-1}$.
By assuming that i) all these blazars have the same mass; ii) the same Doppler boosting and
iii) there is a proportionality between $L_\gamma$ and the accretion luminosity $L_{\rm d}$,
we have suggested that this factor $\sim$100 corresponds to a factor 100 in $L_{\rm d}$.
Assuming further that the greatest $L_\gamma$ occur in systems with $L_{\rm d}/L_{\rm Edd}\sim 1$,
we conclude that the BL Lac phenomenon occurs in systems with accretion disk
radiating at a $L_{\rm d}/L_{\rm Edd}<10^{-2}$ level.
This long chain of arguments may seem somewhat contrived at firsts, but all steps 
can be (and have been) verified by more detailed analysis (and information), and
we can now confirm this important conclusion: 
the accretion rate (in Eddington units) controls if a blazar is a low power and lineless
BL Lac or a powerful and line emitting FSRQ.
Using completely different arguments, Ghisellini \& Celotti (2001) reached the same conclusion
for the two types (FR I and FR II) of radio--galaxies, that are thought to be the parent
population of blazars.

\begin{figure}
\vskip -0.7 cm
\begin{tabular}{l l}
\includegraphics[height=.35\textheight]{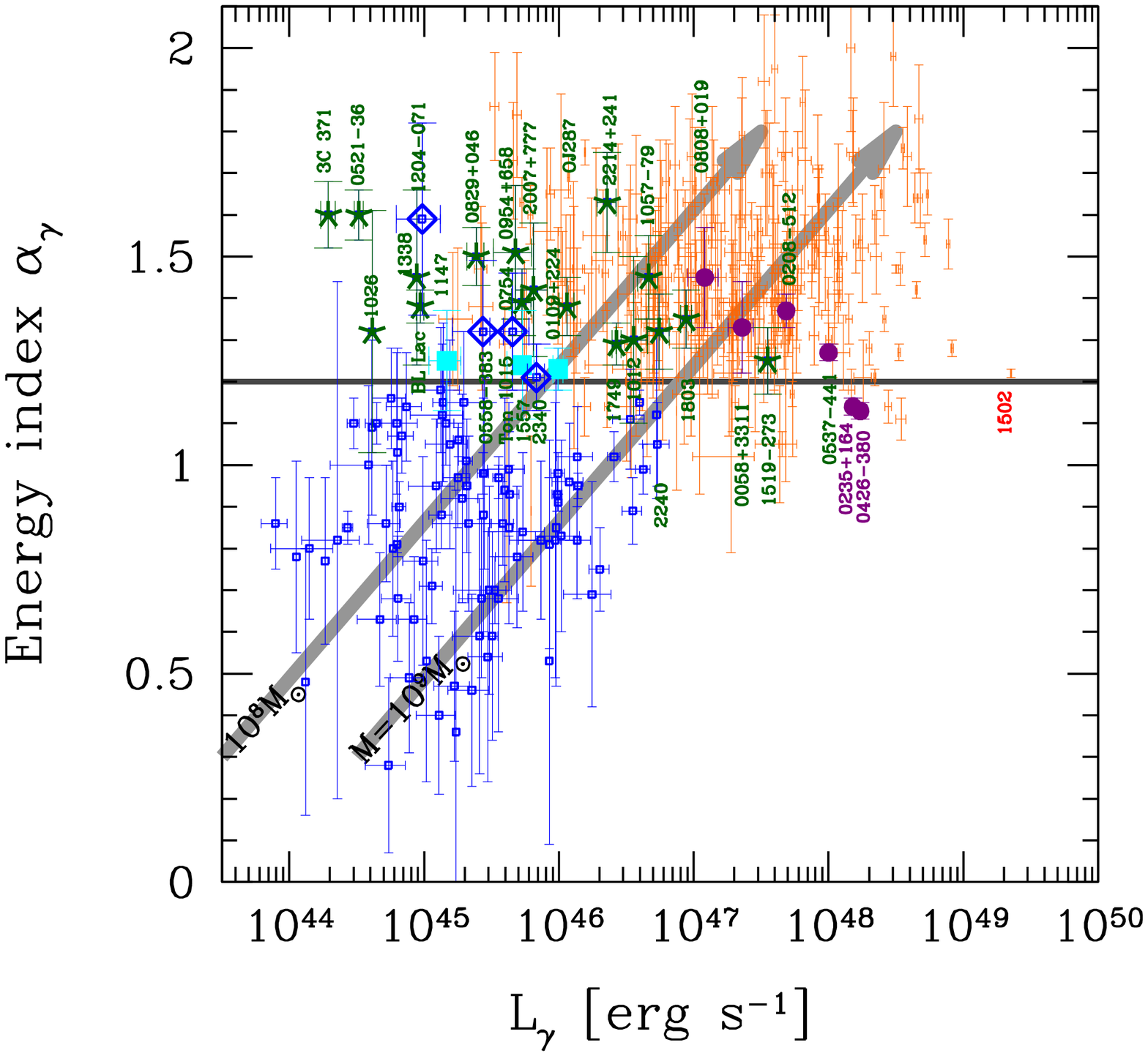}
\includegraphics[height=.35\textheight]{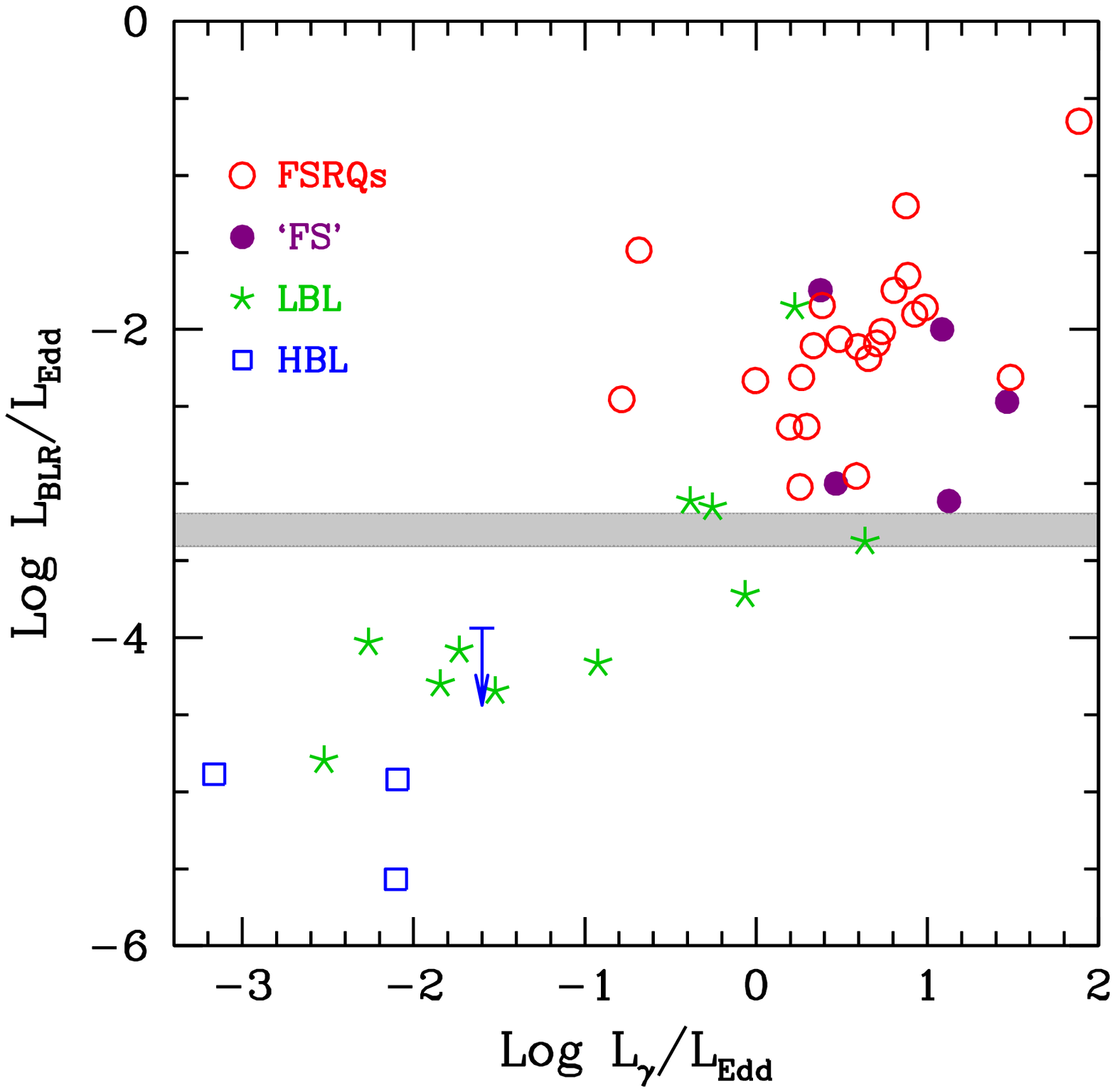}
\end{tabular}
\vskip -0.5 cm
\caption{
{\bf Left:} The energy spectral index $\alpha_\gamma$ as a function
of the $\gamma$--ray luminosity $L_\gamma$ in the band [0.1--10 GeV] 
for all blazars with known redshift present in the 1LAC sample (Abdo et al. 2010a).
Note that some BL Lacs are in the ``FSRQ" region, characterized
by a steep slope ($\alpha_\gamma>1.2$). 
These ``intruders" have been partly re--classified as FSRQs by Ghisellini
et al. (2011) on the basis of the luminosity of their broad emission lines,
measured in Eddington units (filled violet circles).
{\bf Right:} Luminosity of the broad line region (in units
of the Eddington one for sources with at least one broad line 
in their spectrum and with an estimate of the black hole mass)
as a function of the $\gamma$--ray luminosity in units of the Eddington one.
From Ghisellini et al. (2011).
}
\label{divide}
\end{figure}

We may wonder what happens when we start to include, in the $\alpha_\gamma$--$L_\gamma$
plane, the (fainter) blazars detected in the first 11 month of {\it Fermi} life,
with a $>5\sigma$ confidence level.
These are shown in Fig. \ref{divide}.
There is still a ``zone of avoidance" for flat and powerful sources
(although there is still a concern about blazars with no lines and then
no redshift that could populate this zone if their redshift is larger than 2),
but the ``divide" in $L_\gamma$, that was clear in Fig. 1 
of Ghisellini, Maraschi \& Tavecchio (2009), is now absent.
We suggest that this is because of the reasons outlined above: we start to see objects
not perfectly aligned, possibly with smaller black hole masses and then
smaller $L_\gamma$, yet with the same Eddington ratios than before.
Fig. \ref{divide} shows lines of increasing $\dot M/\dot M_{\rm Edd}$ for 
different black hole masses $M$, that can encompass the data points.

\subsection{Different accretion regimes for BL Lacs and FSRQs}

The found ``divide" for BL Lacs and FSRQs (and FR I and FR II radio--galaxies)
can have a simple and straightforward interpretation if there is
a change in the accretion regime, going from radiatively efficient (for
large $\dot M$ and $L_{\rm d}$) to radiatively inefficient (Narayan Garcia \& McClintock 1997).
This change is expected to occur around $L_{\rm d}/L_{\rm Edd} \sim 10^{-2}$,
corresponding to $\dot M/\dot M_{\rm Edd} \sim 0.1$ 
(here $\dot M_{\rm Edd}\equiv L_{\rm Edd}/c^2$, without considering the efficiency of accretion).

When radiatively inefficient, the spectral shape of the disk is very different
form what produced by a standard (i.e. Shakura \& Sunjaev 1973) disk in two respects:
i) the accretion efficiency $\eta$ defined as $L_{\rm d} =\eta \dot M c^2$
becomes a function of $\dot M$, decreasing for decreasing $\dot M$.
Narayan, Garcia \& McClintock (1997) proposed $\eta =\min (0.1;  \dot M/\dot M_{\rm Edd})$.
ii) the fraction of the bolometric luminosity emitted in the optical UV
decreases, so that the ionizing flux decreases even more than linearly
with $\dot M$.
The decreased ionizing flux means that all the broad lines becomes weaker:
we have a BL Lac object.
The decreased amount of seed photons for the IC process implies a weaker
radiative cooling for the relativistic electron in the jet, and they can
thus reach larger typical energies: therefore the two humps peak
at larger frequencies.
The high energy hump is mainly produced by SSC only, and is thus
less dominant (or even weaker) than the synchrotron hump.
These arguments explain the general properties of the blazar sequence.

\subsection{A new classification of blazars}

Traditionally, blazars are classified as BL Lacs or FSRQs according
to the equivalent width (EW) of their broad emission lines (see e.g. Urry \& Padovani 1995).
Objects with a rest frame EW$<$5 \AA\ are called BL Lacs.
This definition has the obvious advantage of being simple and of immediate
use for an observational characterization of the object, and it does
measure the relative importance of the beamed non--thermal continuum 
in the optical band.
On the other hand, EW greater than 5 \AA\ may be the results of a particularly low state 
of the beamed continuum in a source of intrinsically weak lines.
On the opposite side, in several cases a small EW does not imply emission lines of 
low luminosity, being simply the result of a particularly
beamed non--thermal continuum. 
PKS 0208--512 can illustrate this point:
it has an observed MgII emission line of EW$\sim$5 \AA\ (2.5 \AA\ in the rest frame;
Scarpa \& Falomo 1997),  whose luminosity
is close to $10^{44}$ erg s$^{-1}$, stronger than in some FSRQs.
This object is classified as a BL Lac, but all its physical properties are
resembling FSRQs.
When discussing the different physical properties of BL Lacs and FSRQs
it is then confusing having sources like PKS 0208--512 classified as a BL Lac.
We (Ghisellini et al. 2011) have then proposed a new and physically based classification scheme.
It is based on the luminosity of the BLR, measured in Eddington units.
Although this study dealt with a relatively small sample of sources
(all detected by {\it Fermi}), we found that 
$L_{\rm BLR}/L_{\rm Edd}$ correlates with $L_\gamma/L_{\rm Edd}$ (see Fig. \ref{divide})
and with the shape of the SED.
Due to this continuity of properties, it can even become questionable 
the very need to divide blazars into sub--categories.
But if we really want to divide FSRQs from BL Lacs, we propose that 
$L_{\rm BLR}/L_{\rm Edd} \sim 10^{-3}$ (or slightly less) is the dividing value.
This is in very good agreement with what discussed before if the BLR reprocesses 
$\sim$10\% of the accretion disk luminosity:
in this case the dividing disk luminosity is of the order of  $L_{\rm d}/L_{\rm Edd} \sim 10^{-2}$.

\begin{figure}
\vskip -1 cm
\begin{tabular}{l l}
\includegraphics[height=.35\textheight]{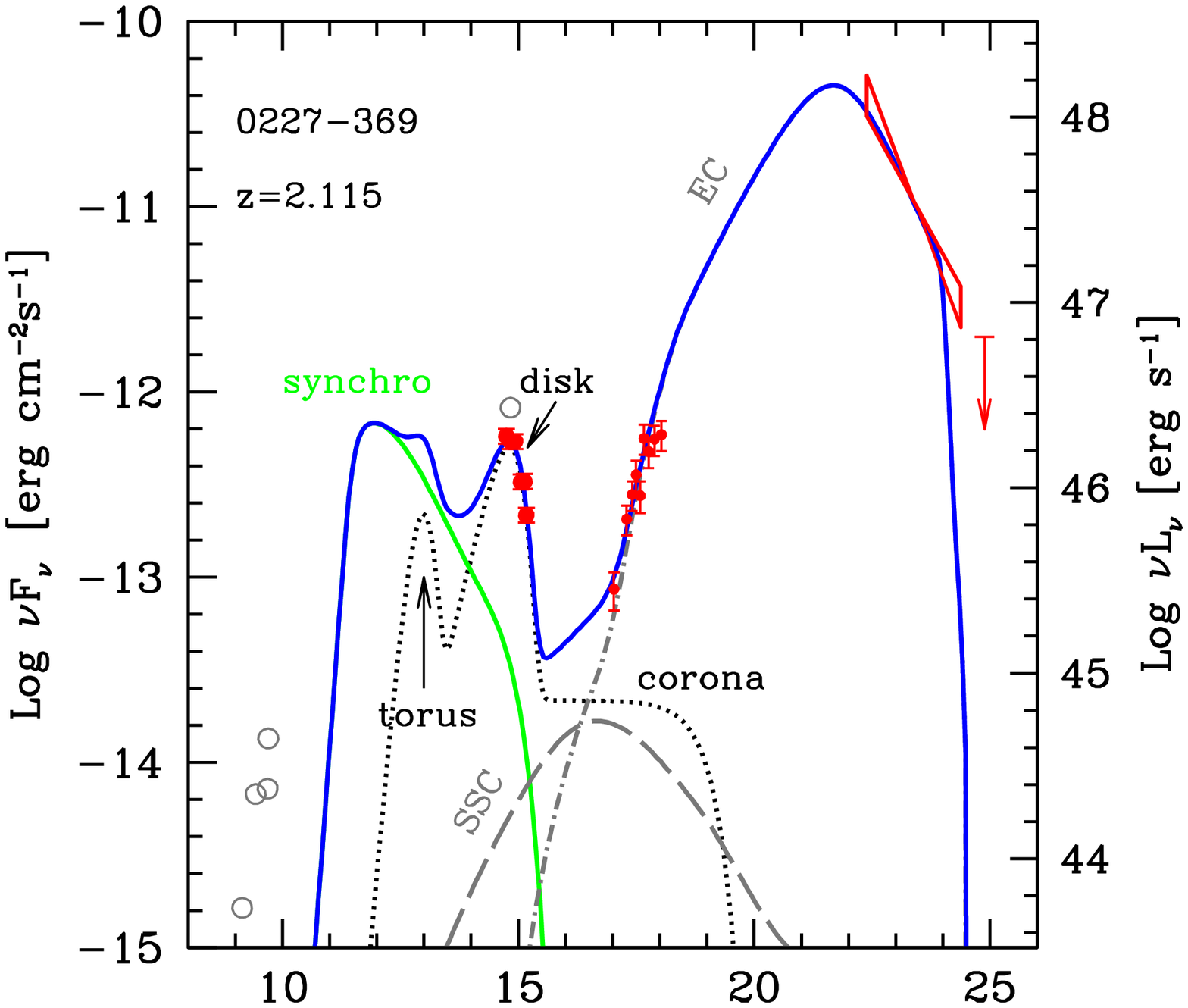}
\includegraphics[height=.35\textheight]{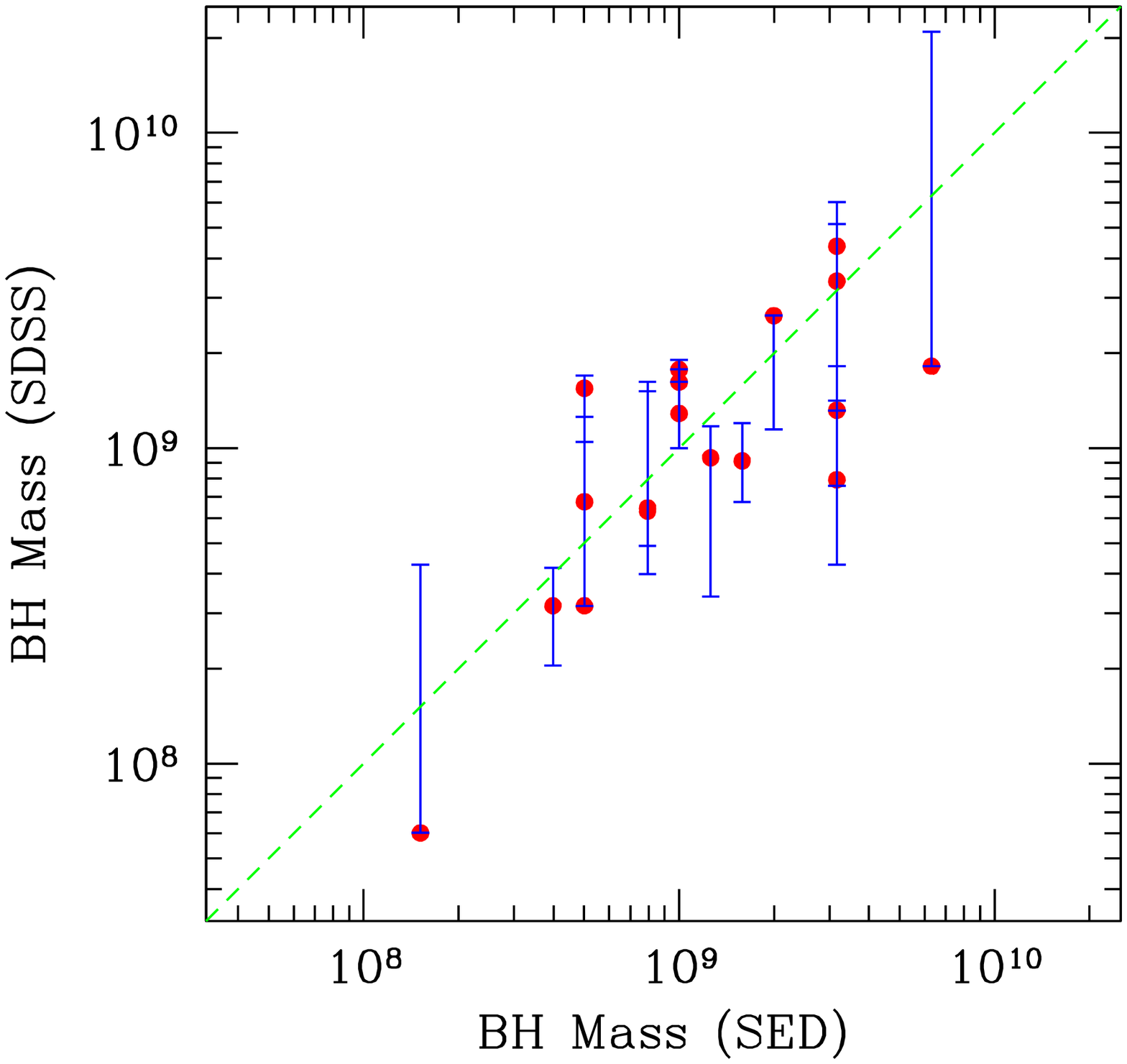}
\end{tabular}
\vskip -0.5 cm
\caption{
{\bf Left:} The SED of the blazar 0227--269 ($z=2.115$) modelled
with the one--zone leptonic model described in Ghisellini \& Tavecchio (2009).
The different components are labelled.
As can be seen, the non thermal jet emission, 
although largely dominating the bolometric output,
leaves the disk emission ``naked". 
{\bf Right:} Black hole masses estimated through the FWHM of the broad emission lines and 
the correlation between the ionizing luminosity and the size of the BLR 
(as calculated for FSRQs in the SDSS DR7 catalog) vs the black hole masses
derived from fitting the optical--UV continuum with
a simple Shakura \& Sunjaev (1973) disk model.
The vertical bars are not error bars, but indicate the range of minimum/maximum black hole
masses estimated using different lines.  
The dashed line indicates equality.
}
\label{mass}
\end{figure}

\section{Black hole masses and accretion}

Fig. \ref{mass} (left panel) shows the SED of the {\it Fermi} detected
FSRQ 0227--369 ($z$=2.115), together with the one--zone synchrotron+IC model 
used to fit the data.
Fig. \ref{mass} shows that in these high redshift powerful blazars 
the non--thermal beamed jet continuum leaves the optical--UV disk
emission unhidden.
By assuming that the emission is produced by a standard Shakura--Sunjaev (1973) 
disk we can estimate both the black hole mass and the accretion rate.
This is the case for the majority of powerful {\it Fermi} detected blazars,
characterized by a relatively steep slope in the GeV energy range and a corresponding
steep synchrotron spectrum above the peak, occurring in 
far IR or sub--mm range.

The main uncertainty on the derived black hole mass ({\it if} the disk van be indeed 
described by a Shakura--Sunjaev model) lies in the amount of absorption 
of the optical--UV data, affected by the intervening Lyman--$\alpha$ systems
along the line of sight (if the blazar are at $z>2$).
In Ghisellini et al. (2010b) we have accounted for this effect, 
estimating the {\it average} number intervening of Lyman--$\alpha$ systems and their optical depth.
The variance around the average can however be large, and this is the main cause
of uncertainty.
On the other hand, this method is in any case competitive with the estimates
derived through the FWHM of the broad emission lines and assuming a relation
between the size of the BLR and the luminosity of the ionizing
continuum (Kaspi et al. 2007; Bentz et al. 2008).
In fact, in Fig. {\ref{mass} (right panel) we compare the black hole masses
estimated by fitting the optical--UV SED with the masses derived through the 
emission lines. 
The vertical bars are not error bars, but indicate the range of masses 
resulting by using different emission lines and/or different scaling 
relations between $R_{\rm BLR}$ and the ionizing luminosity.
The agreement between the masses calculated with the two methods
is reassuring.

The left panels of Fig. \ref{istomass} shows the distribution of black hole masses
for two samples of blazars lying at $z\ge 2$: the left top panel refers to all blazars
detected by {\it Fermi} present in the 1LAC catalog (A10), the bottom panel concerns
all blazars detected by {\it Swift}/BAT (in the 15--55 keV range; Ajello et al. 2009).
Almost all have $M>10^9 M_\odot$, with the ``record holder" blazar 0014+813 
having $M=4\times 10^{10}M_\odot$ (right panel in Fig. \ref{istomass}).
This outrageous large mass comes, formally, from the large IR--optical luminosity,
exceeding $10^{48}$ erg s$^{-1}$, interpreted as accretion luminosity close
to the Eddington limit (Ghisellini et al. 2009). 
Note the luminosity and slope of the IR emission, which is insensitive to absorption.
The derived black hole mass in this blazar is so large to motivate
the search for other explanations.
One hint can come from the not so extraordinary jet luminosity: comparing 
0014+813 with other BAT or LAT detected blazars at high redshift, we find
that in 0014+813 the ratio between the non--thermal and the disk luminosity 
is a factor $\sim$10 less than the other blazars.
This suggests that in this specific object, emitting very close to Eddington,
the accretion disk may not be geometrically thin in its inner part, but it may
have developed a doughnut shape. 
This would make the disk emission not isotropic, but collimated along the normal to the disk.
Viewing the source face on (as we do in this case, since we selected it as a blazar)
we receive a larger than average flux (see Fig. \ref{pjld}: 0014+813 is the source 
with the largest $L_{\rm d}$).

\begin{figure}
\vskip -0.8 cm
\begin{tabular}{l l}
\includegraphics[height=.35\textheight]{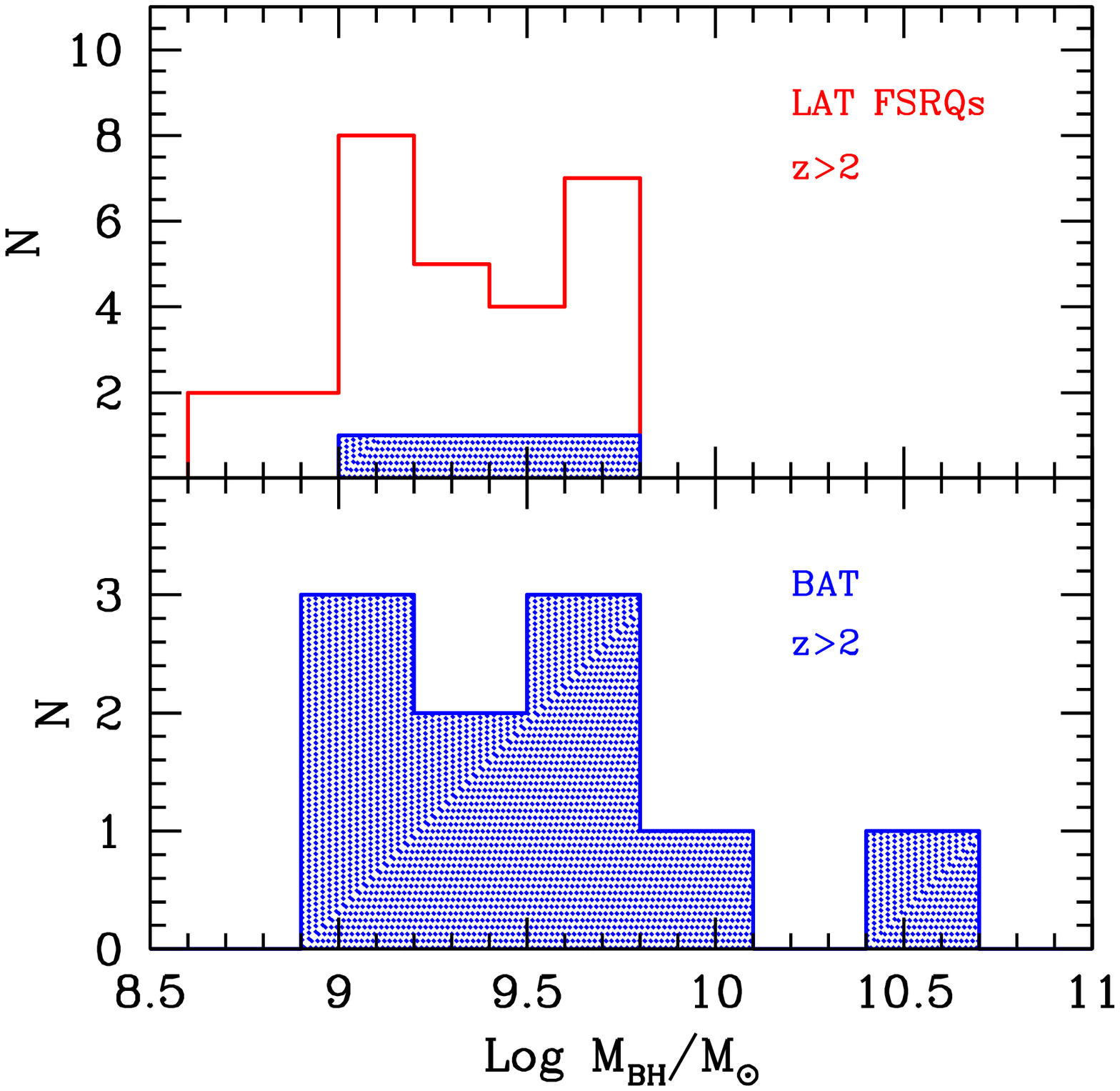}
\includegraphics[height=.35\textheight]{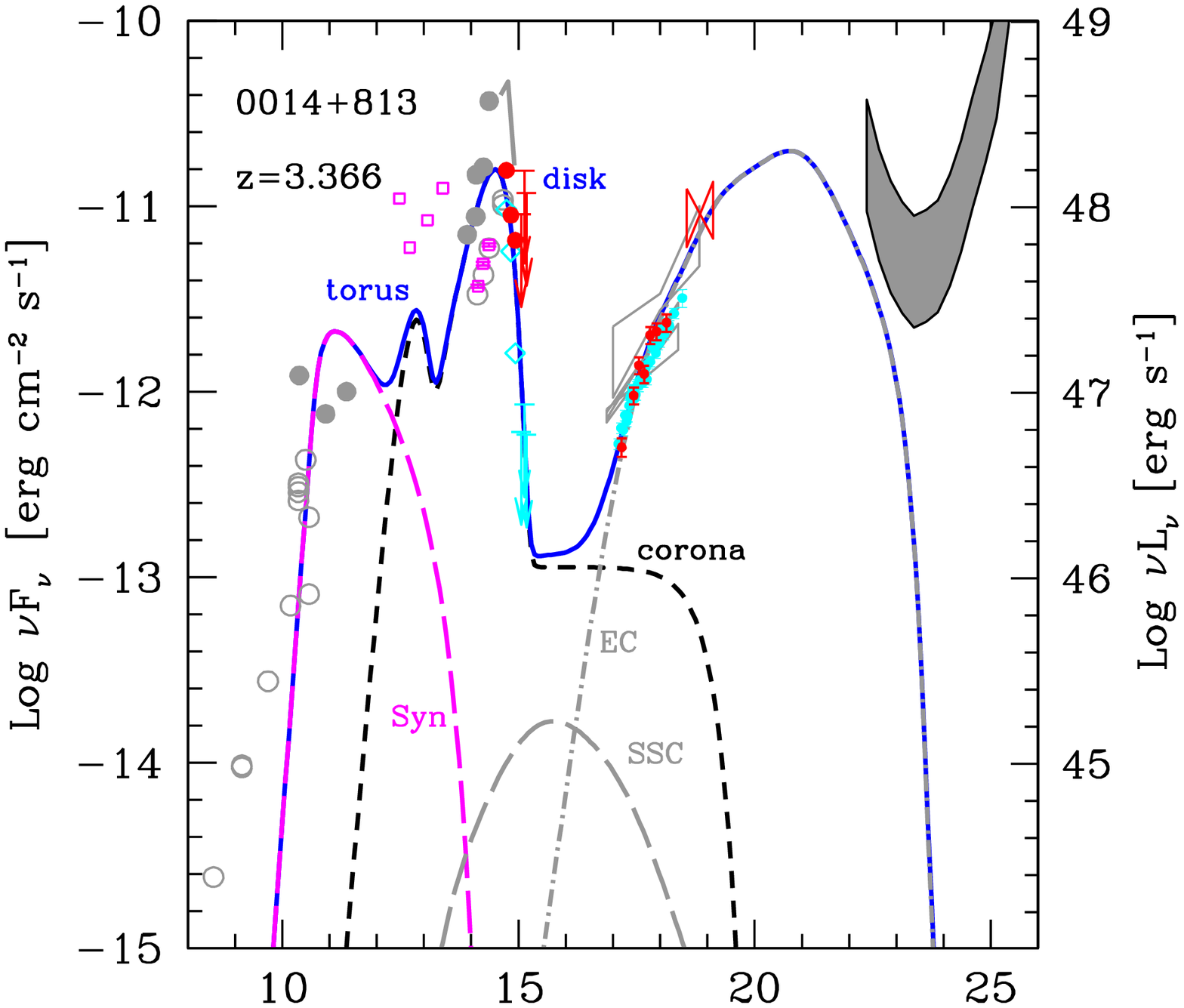}
\end{tabular}
\vskip -0.5 cm
\caption{
{\bf Left:} Black hole mass distribution for blazars at $z>2$ in the {\it Fermi} 1LAC (above)
and {\it Swift}/BAT (bottom) catalogs.
The hatched area in the top histogram corresponds to blazars in common.
Almost all these distant blazars have black holes with masses exceeding $10^9M_\odot$.
{\bf Right:} SED and model of the BAT detected blazar 0014+813.
The black hole mass in this blazar is found to be around 40 billion solar masses,
with an accretion rate close to Eddington. 
Note that the optical luminosity produced by the accretion disk exceeds $10^{48}$ erg s$^{-1}$.
From Ghisellini et al. (2009).
\label{istomass}
}
\end{figure}



\section{Jet powers}

Measuring the jet power is not an easy task, since the radiation we 
see is beamed relativistically, and what we see does not immediately
correspond to what it is produced.
There are however many ways to measure the jet powers:
\begin{itemize}

\item Extended radio emission and/or radio lobes are a sort of calorimeters:
since the radiative cooling times are long for these structures, they transform
the received jet power into magnetic, electron and possibly proton energies.
Knowing the size, and 
using minimum energy (i.e. equipartition) arguments (Burbidge 1959), one can find a lower limit
to the jet power, that heavily depends on the assumed proton energy.
Furthermore, if one can estimate the time needed to form the radio lobe,
one derives the average power received by the lobe, i.e. the average jet power.
This quantity correlates with the luminosity of the narrow emission lines,
that should in turn be a good proxy for the accretion disk luminosity.
It turns out that the average jet power and the disk luminosity calculated
in this way are approximately equal (Rawlings \& Saunders 1991).

\item Detailed X--ray imaging of radio sources revealed the presence of 
``X--ray cavities" (e.g. Allen et al. 2006; Balmaverde et al. 2008) filled instead with 
relativistic radio--emitting electrons. 
One can calculate the $PdV$ work to form these cavities and then guess the
associated jet power.

\item In the VLBI zone we can measure both the size of the emitting knot and
its velocity, often superluminal.
From the radio flux and size one can predict the amount of self--Compton emission,
expected in the X--rays. Comparing with real data one derives a limit to the 
beaming factor. This is turn helps in fixing the required number of relativistic electrons
and magnetic field needed to account for the radio emission we see, 
and thus the kinetic and Poynting flux of the jet.

\item At the scale of hundreds of kpc, some jets have been detected and resolved
in X--rays by the {\it Chandra} satellite (see e.g. Schwartz et al. 2000 for PKS 0637--752).
This emission, interpreted as IC off photons from the microwave background, can
give an estimate the jet power (see Celotti et al. 2001; Tavecchio et al. 2000;
Georganopoulos et al. 2005)

\item The bulk of emission of blazars is emitted in $\gamma$--rays.
This translates in a lower limit to the jet power (it must have more
power than the one radiated). The limit is (see Celotti \& Ghisellini 2008)
\begin{equation}
P_{\rm j} >P_{\rm r} = {L_{\rm obs} \over \Gamma^2}
\end{equation}
It is a lower limit because if the jet spent all its power to produce radiation,
then it would stop.
In powerful sources dominated by the EC process, the emission is anisotropic also 
in the comoving frame, therefore the jet recoils, unless the inertia of the
jet is large.
If the jet were composed only by e$^\pm$, it would decelerate strongly,
so we require the presence of at least one proton every $\sim$10 leptons
(Ghisellini \& Tavecchio 2010). 
They contribute to the jet power. 
Therefore a very robust lower limit to the total jet power
is $P_{\rm jet} > 2P_{\rm  r} \sim 2 L_{\rm obs}/\Gamma^2$.

\end{itemize}

Concerning the latter method, several attempts have been done in the past to find the
jet power and the accretion disk luminosity in blazars and radio--loud objects in general 
(starting from 
Rawlings \& Saunders 1991; 
Celotti et al. 1997; 
Cavaliere \& D'Elia 2002; 
Maraschi \& Tavecchio 2003; 
Padovani et al. 2003; 
Sambruna et al. 2006;
Allen et al. 2006;
Celotti \& Ghisellini 2008;
Ghisellini \& Tavecchio 2008; 
Kataoka et al. 2008;
Ghisellini et al. 2010a (G10);
Bonnoli et al. 2011).
These works found large jet powers, often larger than the luminosity produced by the disk.

The main uncertainty when estimating the power in this way is the contribution
of protons to the bulk kinetic power.
If one assumes one proton for each emitting electron, then the number of
protons depends on the low energy cut--off of the electron distribution
(where most of the electrons are), that in powerful sources is bound to be at
small energies, because the radiative cooling is severe.
In addition, there might be e$^\pm$ pairs, that would limit the required
number of protons and thus lower the total jet power.
But, as mentioned above, the number of pairs cannot be larger than $\sim$10 per proton,
otherwise the  ``Compton rocket" recoil effect becomes too strong.
This is in agreement with 
Sikora \& Madejski (2000) and Celotti \& Ghisellini (2008), who also argued that 
e$^\pm$ pairs cannot be dynamically important, corresponding to a limit of a few
pairs per proton.

\begin{figure}
\vskip -0.8 cm
\begin{tabular}{l l}
\includegraphics[height=.35\textheight]{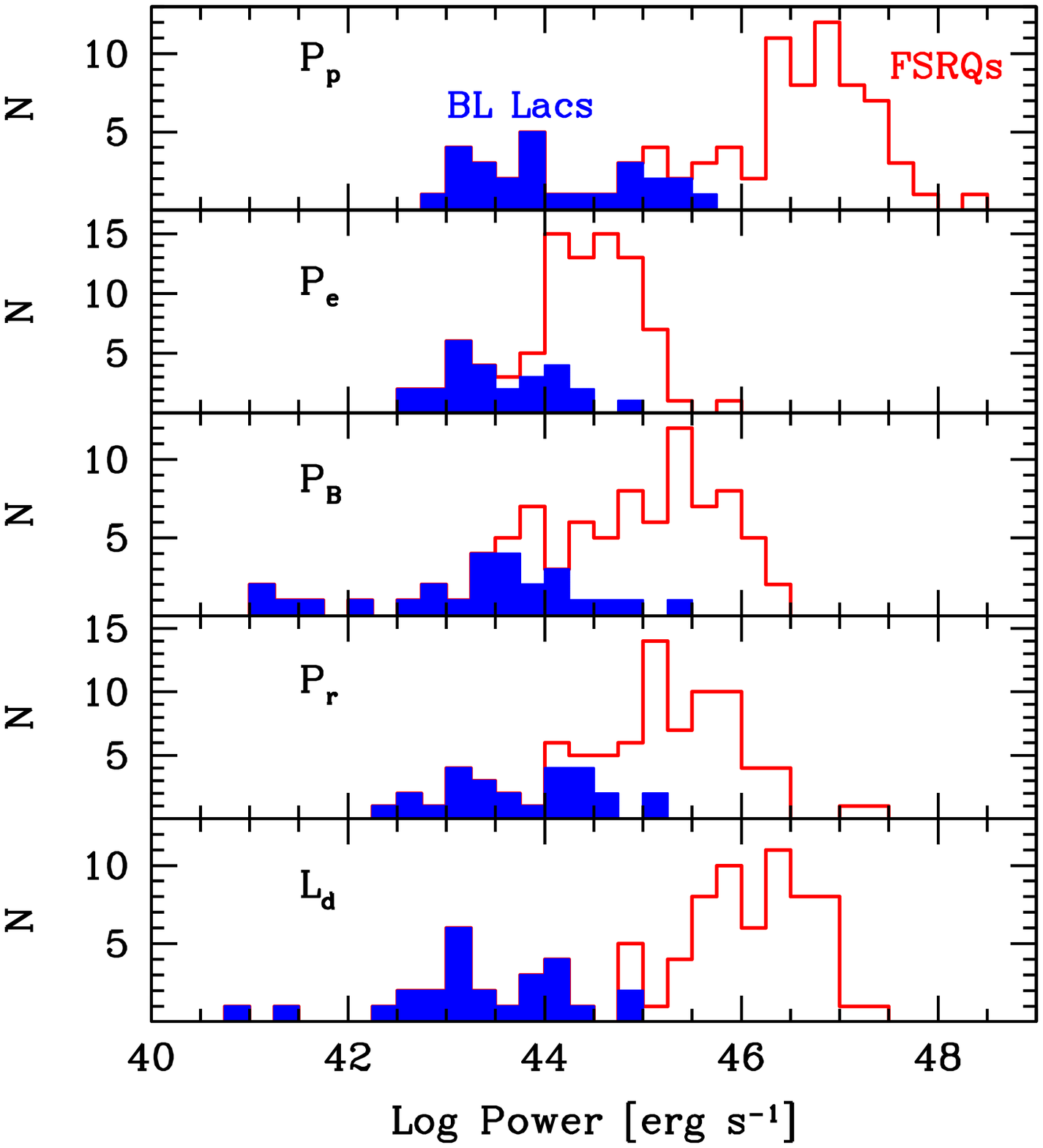}
\includegraphics[height=.35\textheight]{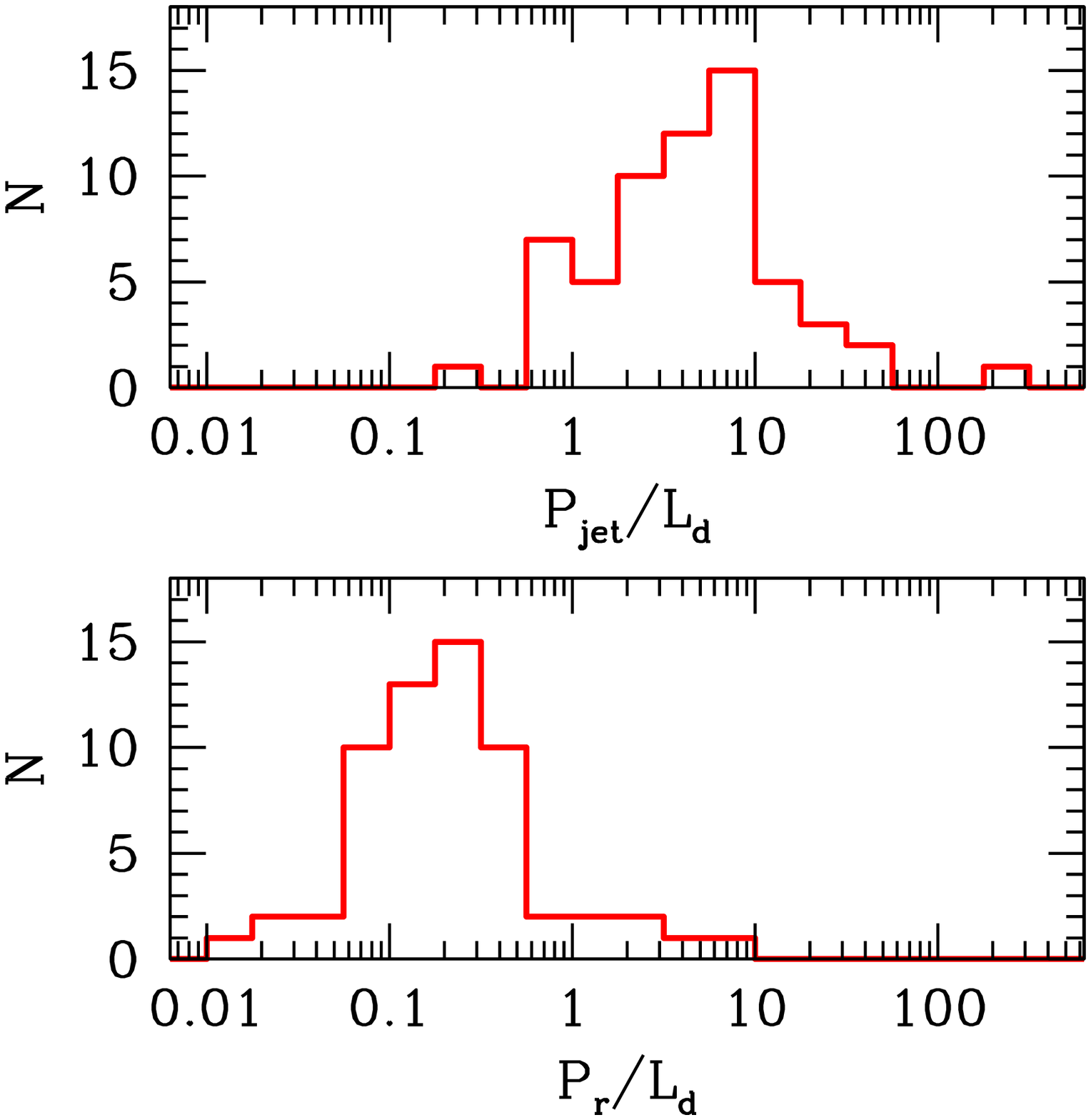}
\end{tabular}
\vskip -0.1 cm
\caption{
{\bf Left:} Distribution of the jet power in different forms (first
4 panels) and of the accretion luminosity (bottom panel).
The blue hatched areas refer to BL Lac object, the red histograms
refer to FSRQs.
The values for the accretion luminosity for BL Lacs are upper limits.
{\bf Right:} Distributions of $P_{\rm jet}/L_{\rm d}$ (top)
and of $P_{\rm r}/L_{\rm d}$ (bottom) for FSRQs only.
From G10.
\label{power}
}
\end{figure}

Fig. \ref{power} shows the histograms of the different forms of power 
carried by the jet.
The shaded areas correspond to BL Lacs.
Besides the power spent to produce the radiation we see ($P_{\rm r}$),
the power is carried in the form of relativistic electrons ($P_{\rm e}$),
magnetic field ($P_{\rm B}$), or cold protons ($P_{\rm p}$).
All these powers can be calculated through:
\begin{equation}
P_{\rm i}\, =\, \pi R^2 \Gamma^2 c U^\prime_{\rm i}
\end{equation}
where $R$ is the size of the emitting blob, assumed to be equal to the cross sectional
radius of the jet, and $U^\prime_{\rm i}$ is the (comoving) energy density of the i$_{\rm th}$
component of the power.

As mentioned, the most robust, almost model--independent, lower limit to the 
jet power is $P_{\rm r}$, spent by the jet to produce its radiation. 
For FSRQs, the distribution of $P_{\rm r}$ extends to larger values than
the distribution of $P_{\rm e}$.
The distribution of $P_{\rm B}$ is at slightly smaller values than the 
distribution of $P_{\rm r}$, indicating that the Poynting flux cannot
be at the origin of the radiation we see.
As described in Celotti \& Ghisellini (2008), this is a direct consequence
of the large values of the Compton dominance (i.e. the ratio of 
the Compton to the synchrotron luminosity is small), since
this limits the value of the magnetic field.

To justify the power that the jet carries in radiation we are forced to consider 
protons.
If there is one proton per electron (i.e. no pairs), then
$P_{\rm p}$ for FSRQs is a factor $\sim$10--100 larger than $P_{\rm r}$,
meaning an efficiency of 1--10\% for the jet to convert its bulk kinetic
motion into radiation (see the right panel of Fig. \ref{power})
This is reasonable: most of the jet power in FSRQs goes to form and energize the 
large radio structures, and not into radiation.
                       
We then conclude that {\it jets should be matter dominated}, at least at the 
scale (hundreds of Schwarzschild radii from the black hole) 
where most of their luminosity is produced. 
The bottom left panel of Fig. \ref{power} shows the distribution of the 
disk luminosities.
In this case the shaded area corresponds to upper limit for BL Lac objects,
and not to actual values.
This $L_{\rm d}$ distribution lies at intermediate values between 
$P_{\rm r}$ and $P_{\rm p}$.

\begin{figure}
\vskip -0.5 cm
\begin{tabular}{l l}
\includegraphics[height=.35\textheight]{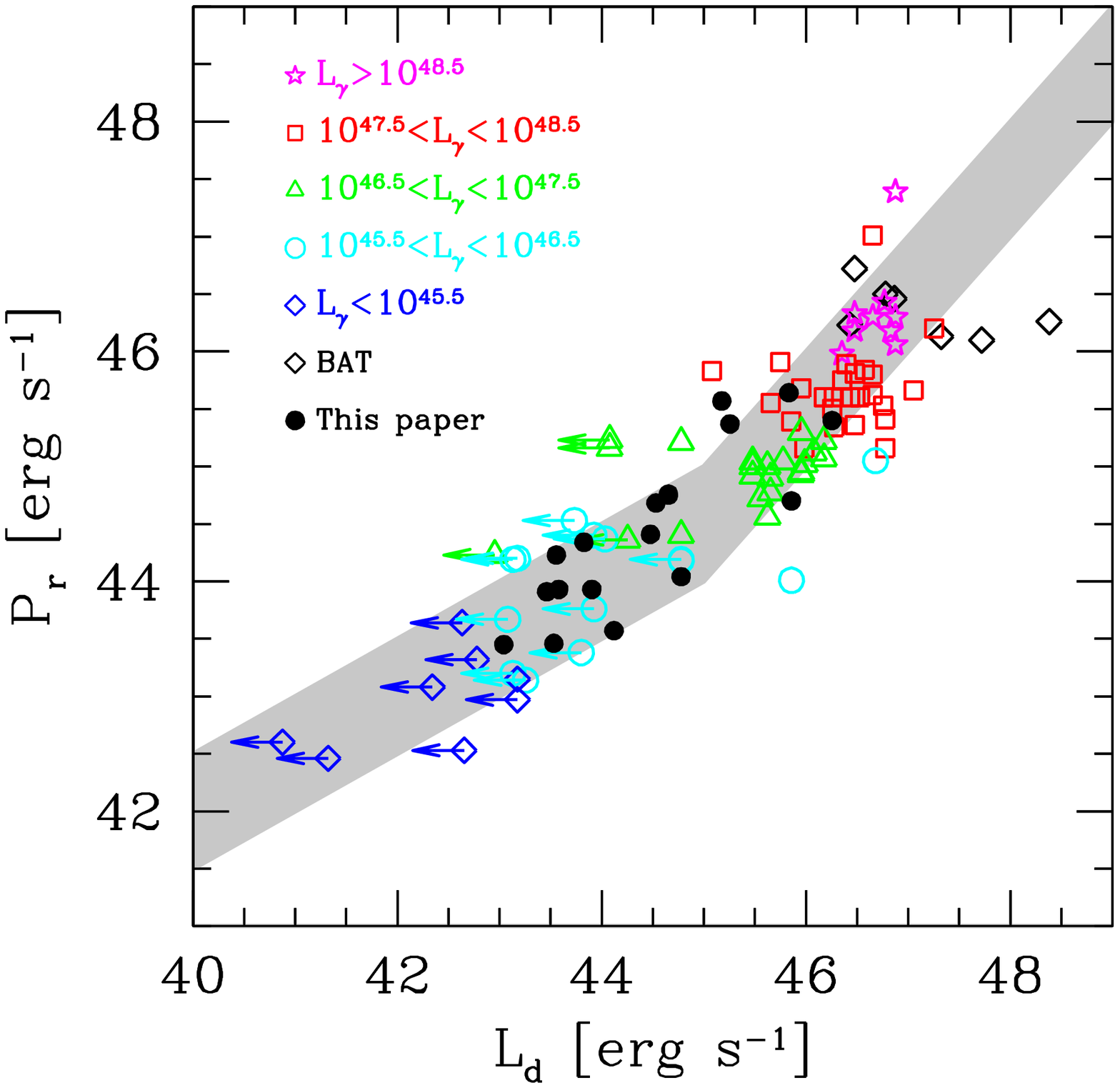}
\includegraphics[height=.35\textheight]{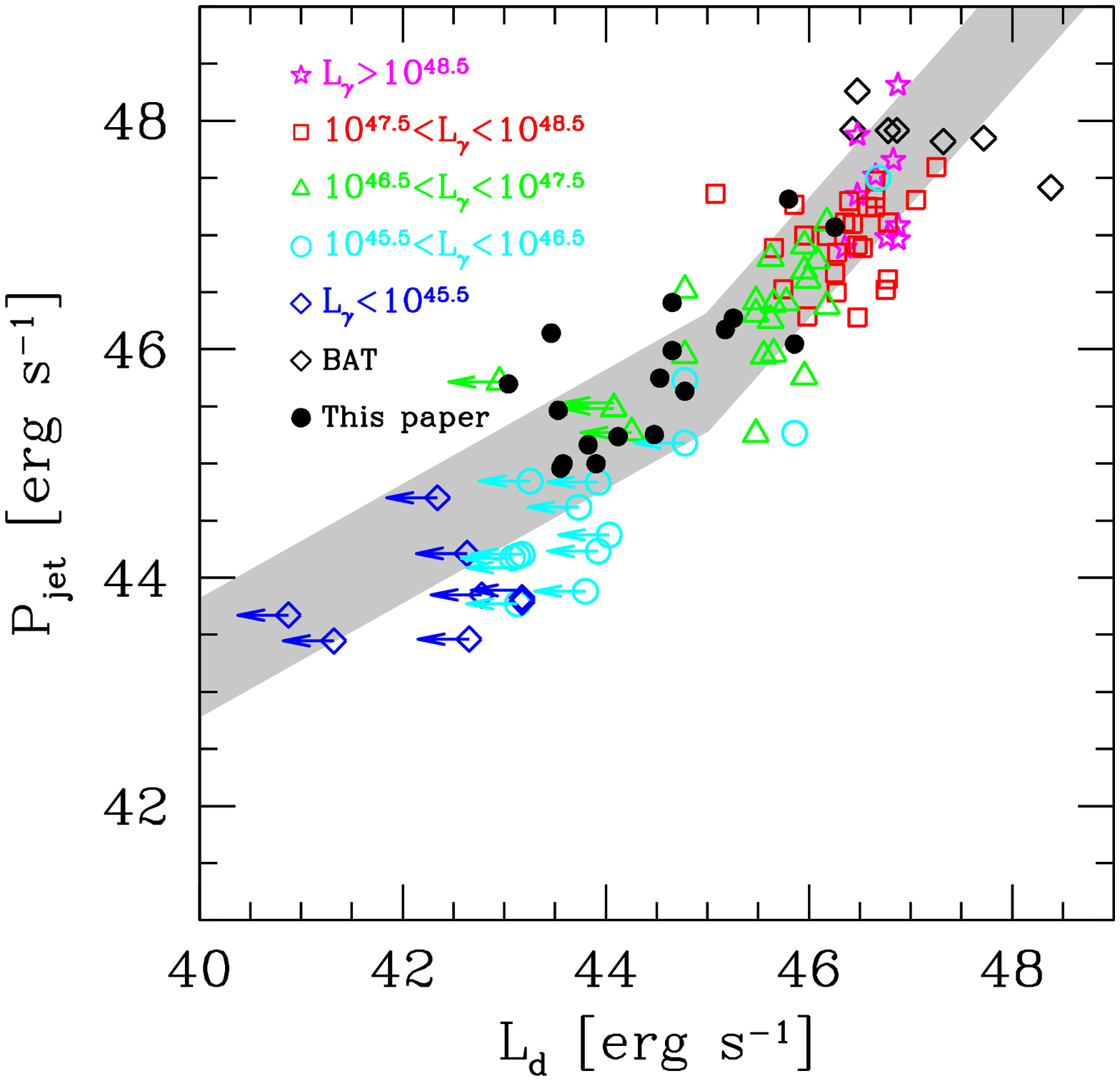}
\end{tabular}
\vskip -0.3cm
\caption{
{\bf Left:} Power of the jet spent in the for of radiation $P_{\rm r}$
as a function of the accretion luminosity $L_{\rm d}$.
Black symbols are the estimates in Ghisellini et al. (2011), BAT points (grey diamonds)
come from the high--redshift blazars present in
the 3 year all sky survey of BAT (and studied in Ghisellini et al. 2010a), the other points and
upper limits come from G10, and are divided according to their $\gamma$--ray
luminosities, as labelled.
{\bf Right:} total jet power $P_{\rm jet}$ as a function of $L_{\rm d}$. 
One proton per emitting electron is assumed.
The grey stripes indicates what expected if $P_{\rm jet}\propto P_{\rm d}\propto \dot M$ for all
luminosities, while $L_{\rm d}\propto \dot M^2$ and $L_{\rm d} \propto \dot M$ at low 
and high values of $\dot M$, respectively. 
Therefore, for low $L_{\rm d}$, we expect $P_{\rm jet}\propto P_{\rm r}\propto  L_{\rm d}^{1/2}$,
while, for large $L_{\rm d}$,  we expect $P_{\rm jet}\propto P_{\rm r}\propto  L_{\rm d}$.
\label{pjld}
}
\end{figure}

\section{The jet--disk connection}

Fig. \ref{pjld} shows $P_{\rm r}$ (left) and 
$P_{\rm jet}\equiv P_{\rm p}+P_{\rm e}+P_{\rm B}$ (right) as
a function of the thermal disk luminosity $L_{\rm d}$.
Arrows correspond to BL Lacs for which only an upper limit on $L_{\rm d}$ 
could be derived.
The different symbols correspond to blazars of different
$\gamma$--ray luminosities, and one can see that $L_\gamma$
correlates both with $P_{\rm jet}$ and $L_{\rm d}$.
Furthermore we also show blazars not yet detected by {\it Fermi}, but 
present in the all sky survey of {\it Swift}/BAT, and lying at $z>2$.
As discussed in G10, there is a significant correlation between
$P_{\rm jet}$ and $L_{\rm d}$ for FSRQs, which remains highly significant
even when excluding the common redshift dependence.
The slope of this correlation for FSRQs is consistent with being linear,
with $P_{\rm jet}$ slightly larger than $L_{\rm d}$.
For BL Lacs $P_{\rm jet}\gg L_{\rm d}$, but for the 
majority of BL Lacs we can estimate only an upper limit on $L_{\rm d}$,
so the slope of the relation (if any) $P_{\rm jet}$--$L_{\rm d}$ is not known.

The grey stripe is indicative of what expected if $P_{\rm jet}$ 
always traces $\dot M$ (irrespective of the accretion regime, ADAF or radiatively efficient), 
while $L_{\rm d} \propto \dot M^2$ at low luminosity, and $L_{\rm d} \propto \dot M$ 
above some critical value, marking the passage from radiatively inefficient to efficient disk.
If true, this implies that
\begin{eqnarray}
P_{\rm jet} & = & k L_{\rm d},  \qquad \qquad L_{\rm d}>L_{\rm c} \nonumber \\
P_{\rm jet} & = & k \left( L_{\rm d} L_{\rm c}\right)^{1/2}, \quad L_{\rm d}<L_{\rm c}  
\end{eqnarray}
where $k$ should be of order of (but somewhat larger than) unity.
The critical luminosity should be of the order of $L_{\rm c}\sim 10^{-2} L_{\rm Edd}$,
corresponding to $\dot M_{\rm c}\sim 0.1 \dot M_{\rm Edd}$.

Limiting ourselves to FSRQs, we can conclude that the jet power is proportional to 
the mass accretion rate, but it can be even greater than the accretion luminosity.
3C 454.3, during major flares, showed $P_{\rm d}>L_{\rm d}$, and note that $P_{\rm r}$
is a very robust {\it lower limit} to $P_{\rm jet}$.
The proportionality between $P_{\rm jet}$ and $L_{\rm d}$ call for an important role of accretion in
powering jets, but the fact that $P_{\rm jet}$ can be larger than the disk luminosity
makes this possibility very unlikely.
This paradox can be solved in two ways.

\vskip 0.3 cm
\noindent
{\bf Jets powered by accretion only ---} 
Jolley et al. (2009, see also Jolley \& Kuncic 2008), propose that,
in jetted sources, a sizable fraction of the accretion power
goes to power the jet.
As a result, the remaining power for the disk
luminosity is less than usually estimated
by setting the efficiency $\eta\sim0.08$--0.1.
This implies a larger mass accretion rate to sustain $L_{\rm d}$.
The total gravitational energy produced by the mass accretion rate, 
with total efficiency $\eta=\eta_{\rm d}+\eta_{\rm jet}$ goes only in part to produce 
the disk luminosity (with efficiency $\eta_{\rm d}$), while the
rest (with efficiency $\eta_{\rm jet}$) goes to power the jet.
Our results would require $\eta_{\rm jet}>\eta_{\rm d}$.

\vskip 0.3 cm
\noindent
{\bf Jets powered by the black hole spin ---} 
The rotational energy of a maximally spinning black hole is large:
$0.29 Mc^2$ (i.e. $5\times 10^{62} M_9$ erg).
However, this huge reservoir of energy,
amply sufficient to power a strong jet for its entire lifetime,
must be extracted at a sufficiently rapid pace.
The observed relation between $L_{\rm d}$ and $P_{\rm jet}$
can be explained by linking the extraction 
of the hole rotational energy to the accretion process.
The idea is that, as in Blandford \& Znajek (1977) process,
there must be a magnetic field to tap the rotational hole energy.
This magnetic field must be produced by the disk, therefore 
by accretion.
In this case the accretion rate enters because it produces the catalyzer
of the process.
The maximum magnetic field that a disk can sustain will have
an energy density of the order of the gravitational energy
of the matter (if larger, it would disrupt the disk).
Therefore  $B^2/(8\pi) \sim \rho v_\psi^2$ ($v_\psi$ is the
Keplerian velocity of the matter in the disk).
The total efficiency of the process has been
debated in recent years 
(e.g. Moderski \& Sikora 1996; Ghosh \& Abramowicz 1997; 
Livio, Ogilvie \& Pringle 1999; McKinney 2005; Garofalo 2009;
Krolik \& Hawley 2002), and is not yet clear if 
the Blandford \& Znajek process can indeed account for the
huge powers that $\gamma$--loud jets are demanding.

\section{Summary and conclusions}

Understanding extragalactic relativistic jets is not an easy task:
after almost half a century since their discovery we still do not know
what produces and collimates them.
On the other hand our knowledge of jets has steadily increased over the years,
with the most recent advances obtained through $\gamma$--ray observations,
disclosing the jet emission where it peaks.
Besides the high energy observations, the other key quantities that
start to be known for several relativistic jets is the mass of the black hole and the 
mass accretion rate of their disks.
This allows to measure quantities in Eddington units and to compare the
jet powers with the accretion luminosities.
The key results are:
\begin{itemize}

\item The radiative jet power, $P_{\rm r} =L_{\rm bol}/\Gamma^2$
is a lower limit of the jet power, and is model--independent (apart
from the value of $\Gamma$).

\item Jets cannot be magnetically dominated, at least at the scale
where they produce most of the emission ($\sim10^3$ \sc\ radii).
If there is one proton per emitting particle, then $P_{\rm jet}
\sim$(10--100)$\times P_{\rm r}$.

\item In powerful FSRQs, the optical--UV flux can be produced by the
accretion disk. In these blazars we can then measure the black hole mass
and the accretion rates. 
The comparison of the mass inferred by fitting a standard disk spectrum
with the mass derived by other methods is reassuring.

\item Jets are produced for all accretion rates measured in Eddington units.

\item The different ``look" of BL Lacs and FSRQs could be the result
of a different environment, in turn dependent on the accretion regime of the
accretion disk. This leads to a new classification scheme,
dividing BL Lacs and FSRQs on the basis of the value of $L_{\rm BLR}/L_{\rm Edd}$
smaller or greater than $\sim10^{-3}$ 
(corresponding roughly to $L_{\rm d}/L_{\rm Edd}\sim 10^{-2}$).

\item The jet power correlates with the luminosity of the accretion.

\begin{figure}
\begin{center}
\includegraphics[height=.4\textheight]{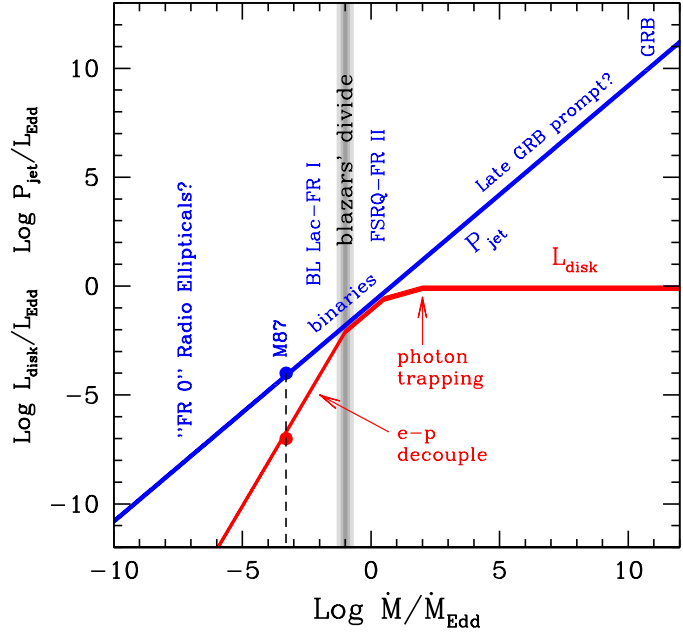}
\end{center}
\caption{
Sketch illustrating $P_{\rm jet}$ and $L_{\rm d}$ as
a function of $\dot M/\dot M_{\rm Edd}$.
It is assumed that the jet power always scales linearly with $\dot M$,
while accretion rates below a critical value produce radiatively inefficient
accretion disks. 
In this case the object looks like a BL Lac (if aligned) or a FR I (if misaligned).
The grey stripe indicates the critical $\dot M \dot M_{\rm Edd}\sim 0.1$, 
producing the blazars' divide at $L_{\rm d}/L_{\rm Edd}\sim 10^{-2}$ equivalent
to $\dot M/\dot M_{\rm Edd}=0.1$.
At very high $\dot M/\dot M_{\rm Edd}$ we find GRBs.
Apart form their proper prompt emission, in which they emit $\gamma$--rays,
there is a phase (lasting a few hours) of X--ray emission (with flares super--imposed)
that has been interpreted as due to the central engine, but accreting a smaller
(and decreasing rate). This phase is labeled ``Late GRB prompt?".
The ``FR 0" radio ellipticals are a new population of radio sources (Baldi et al. 2009)
having the same core radio luminosity of FR Is, but hundreds of times less power in the
extended emission.
\label{final}
}
\end{figure}

\item Fig. \ref{final} shows schematically how the power of jets and
the accretion luminosity could be related to $\dot M$ in Eddington units.
This allows to properly compare objects with different black hole masses,
such as, besides radio--loud AGNs, Galactic X--ray binaries and Gamma Ray Bursts.
The jet power could be {\it always} proportional to $\dot M$.
The disk luminosity, instead, is reduced at both ends of $\dot M/\dot M_{\rm Edd}$,
by the decoupling between electron and protons (at the low end) and by 
photon trapping (at the high end).
Blazars (and radio--galaxies) have a continuity in the power of their jets,
but the different accretion regimes create a discontinuity in environment:
above $\dot M/\dot M_{\rm Edd}=0.1$ (equivalent to $L/L_{\rm Edd}=10^{-2}$)
the disk is ``standard", photoionizes the clouds of the BLR and can illuminate
a dusty IR torus.
The environment is therefore rich in seed photons that the jet emitting region can scatter
to high energies. 
The radiative cooling is severe, the typical electron energies are relatively small,
and the overall SED peaks in the far IR and in the MeV bands.
Below $\dot M/\dot M_{\rm Edd}=0.1$ the disk becomes radiatively inefficient and
the photoionizing flux is largely reduced, the emission lines becomes weak or absent.
Radiative cooling is less severe, electrons can reach large energies, producing
a SED peaking in the UV (or X--ray) and in the GeV (or TeV) bands.
The critical, dividing, value of $\dot M/\dot M_{\rm Edd}$ translates into the
``divide" between BL Lacs and FSRQs.
Fig. \ref{final} shows also the location of other class of jetted sources in the same
plane. 
Galactic binaries, when producing a jet, should lie in the same region of blazars,
while Gamma Ray Bursts, having $\sim10^{12}$ Eddington luminosities, are at the extreme
top right of the plane.
The figure reports also the location of M87, a nearby radio--galaxy (TeV emitting)
thought to be a misaligned blazar, with a radiatively inefficient accretion disk.
Finally, Fig. \ref{final} shows the predicted location of the newly 
discovered population of weak radio--galaxies (Baldi et al. 2009)
with core radio luminosities similar to FR I sources, but with an extended
emission hundreds of times weaker, that we call ``FR 0".

\item Blazars are powerful, and therefore we can detect them up
to large distances.
Moreover, in powerful blazars the non--thermal SED leaves unhidden
the disk emission, allowing us to measure their black hole mass.
As a consequence {\it we can find heavy black holes at large redshifts}.
For each detected blazar, there must be
a few hundreds of similar sources pointing in other directions.
These simple considerations imply that the search for young (i.e. high redshift)
and heavy (i.e. $M>10^9M_\odot$) black holes can largely benefit from blazars.
Limits on the mass function (for $M>10^9M_\odot$) at $z\gsim4$ derived from the radio loud
population are comparable (Volonteri et al. 2011)
with those derived from the SDSS (radio--quiet) quasars (i.e. Hopkins et al. 2007).

\end{itemize}

It can be instructive to conclude this review by pointing out some of the problematic
issues that could be solved in the near future (this is a subjective choice...).
\begin{itemize}

\item {\it One or two (or more) subclasses of blazars? ---}
We have up to now divided blazars into BL Lacs and FSRQs, further subdividing BL Lacs 
according to where their SED peaks (HBL, LBL, IBL, Padovani \& Giommi 1995) 
and FSRQs according if they are lobe or core dominated or if 
they have a high level of optical polarization or not (HPQ, LPQ).
But the new results coming from $\gamma$--rays suggest
that the key quantities, besides orientation, controlling how these
subclasses of radio--loud sources look like are the power of their jets
and the strength of their accretion disks.
It may be more productive to think to radio--sources as a single population,
that appears different, but has instead the same basic jet structure.
One example: for many years we discussed the cosmic evolution properties
of BL Lacs and FSRQs separately, while now a unified approach may be more fruitful
(in line with Maraschi \& Rovetti 1994; Cavaliere \& D'Elia 2002).

\item {\it A one--zone simple model? ---} 
A single--zone model is certainly a crude simplification.
After all we do see several knots in the radio jets, so we do know that
some emission come from different structures in the jet.
At the same time, this simplification allowed a huge step in our
understanding of the basic physics of the jet emission.
What remains to be done it to asses the importance of additional 
emission regions. Can they become sometimes dominant? In what
frequency ranges? Are they really necessary to understand the 
physics of jets or are they ``second order" effects?
An example: the detection of radio--galaxies both by {\it Fermi}
(Abdo et al. 2010b) and by Cherenkov telescopes (e.g. 
M87: Aharonian et al. 2004; CenA: Aharonian et al. 2009; NGC 1275: Mariotti et al. 2010;
IC 310: Alecksic et al. 2010)
would be difficult to understand by assuming an emission
region moving with a very large bulk Lorentz factor, and points to
a more structured jet (e.g. a decelerating flow as in Georganopoulos \& Kazanas 2003;
a spine/layer as in Ghisellini et al. 2005and in
Tavecchio \& Ghisellini 2008 or a reconnection region as Giannios et al. 2009b).

\item {\it Ultrafast TeV variability ---}
This is related to the previous issue, as it must be produced by a very small region,
possibly moving with a very large $\Gamma$.
The novelty in this field is the detection of 10 minutes flux variation
in a FSRQs with broad lines (in 1222+216, Aleksic et al. 2011).
So this phenomenon occurs not only in ``classical" low power TeV BL Lacs,
but also in powerful sources (see Tavecchio et al. 2011, in prep, for
a discussion of possible models).

\item {\it Relativistic jets in Galactic binaries ---}
Are they a scaled down version of extragalactic jets?
If so, there is a lot to learn from these sources, because
they can give us a ``movie" of the entire lifetime of the jet
(i.e. one year in the life of a jetted Galactic binary is equivalent to
$\sim 10^8$ years of a jetted AGN).
The recent detection of Cyg X--3 by {\it AGILE} and {\it Fermi} 
(Tavani et al. 2009; Corbel et al. 2010) is intriguing.

\item {\it Role of the black hole spin ---}
This is a long standing issue, so it should be prudent to think
to a solution of this issue not in the immediate future, but on a longer timescale.
On the other hand there have been in the recent literature many studies
aiming to estimate the black hole spin in jetted sources
(among others: Fender et al. 2010; Daly 2009; McClintock et al. 2010).
While not conclusive, they are the first attempts to solve this issue
experimentally (i.e. observationally), and this line of research
could improve considerably.

\end{itemize}


{\bf Acknowledgments}

It is a pleasure to thank Annalisa Celotti, Luigi Foschini, Laura Maraschi and Fabrizio Tavecchio 
for years of fruitful collaboration.

\end{document}